\documentclass[preprint,preprintnumbers,amsmath,amssymb,showpacs,superscriptaddress]{revtex4}
\usepackage{graphicx}
\usepackage{dcolumn}
\usepackage{bm}

\begin{document}

\title{Testing hadronic interaction packages at cosmic ray energies}
\author{C. A. Garc\'ia Canal}
\affiliation{Departamento de F\'isica, Universidad Nacional de La
Plata,  C.C. 67-1900 La Plata, Argentina} \affiliation{IFLP
(CONICET), Universidad Nacional de La Plata,  C.C. 67-1900 La
Plata, Argentina}
\author{S. J. Sciutto}
\affiliation{Departamento de F\'isica, Universidad Nacional de La
Plata,  C.C. 67-1900 La Plata, Argentina}
\affiliation{IFLP
(CONICET), Universidad Nacional de La Plata,  C.C. 67-1900 La
Plata, Argentina}
\author{T. Tarutina}
\email{tarutina@fisica.unlp.edu.ar} \affiliation{Departamento de
F\'isica, Universidad Nacional de La Plata,  C.C. 67-1900 La
Plata, Argentina}

\date{\today}

\begin{abstract}
A comparative analysis of the secondary particles output of the
main hadronic interaction packages used in simulations of
extensive air showers is presented. Special attention is given to
the study of events with very energetic leading secondary particles,
including diffractive interactions.
\end{abstract}


\maketitle

\section{Introduction}

Cosmic ray physics relies strongly on the simulations of
air showers \cite{Kna03}. Information on the primary particle energy and mass
comes from a hadronic interaction that provides
secondary particles with their masses, multiplicities and energies which,
unknown in principle, are very difficult to model accurately.

Meanwhile, hard hadronic interaction are well described in the
framework of QCD, the soft hadronic interaction observables cannot
be calculated from  first principles and a combination  of
 empirical parameterizations and fundamental theoretical ideas is
used to model them. The parameterization constitutes another type
of difficulty as the accelerator data are available for much lower
energy, another kinematic region and different projectile-target
configuration.

In this work we comparatively analyze the hadronic interaction
models for very high primary energies. We consider the models
included in the packages: SIBYLL2.1 \cite{Fle94,Eng99}, QGSJET01c
\cite{Kal97}, QGSJETII-3 \cite{Ost06}, and EPOS1.6 \cite{Wer06},
which is the successor of NEXUS \cite{Dre01}.
There is an important difference in the way that different models
describe hadronic interaction data. In the next section a brief
description of these packages will be given.

Because of their importance for cosmic ray shower development, we pay
special attention to those particular events characterized by a small
number of secondaries that include a leading particle carrying
a substantial fraction of the projectile energy. We will call them
Very Energetic Leading Particle (VELP) events. While a precise definition of VELP
events is presented in section III, we can say that most of such
events correspond to diffractive processes.
 From the
theoretical point of view, a diffractive process is a high energy
hadronic reaction where no quantum numbers are exchanged between
the colliding particles \cite{Bar02}. From the point of view of
shower development those processes are effective as a way to
transport the primary energy deep into the atmosphere, thus
influencing the position of the shower maximum $X_{max}$. This observable is
one of the most important parameters in Extensive Air Shower (EAS)
physics and is used to
deduce the chemical composition of primary cosmic rays.

In Ref.\cite{Lun04} the three hadronic interaction packages
SIBYLL2.1, QGSJET01c and DPMJET were extensively compared. First,
the observables of individual collisions were studied and then the
shower development was simulated using  SIBYLL2.1 and QGSJET01c.
It was found that the relative probability of diffractive
processes
 during the shower development has a non-negligible influence over
the longitudinal profile as well as the distribution of muons at
ground level. Since that time, new packages of hadronic
interactions have been released, namely QGSJETII-3 and EPOS1.6.
Presently, EPOS is widely used in EAS simulations. The appearance
of these models motivated us to accomplish a thorough systematic
study and comparison of these packages. In this paper, we present
the observables generated by SIBYLL2.1, QGSJETII-3 and EPOS1.6 and
we also include the results of QGSJET01 for comparison.

This paper is organized as follows: in the next section
we review the main features of the different interaction packages; in
section III the details of the performed calculations are presented
and the results are discussed; section IV contains a summary and
our conclusions.

\section{Hadronic interaction models}

At intermediate energies soft processes dominate hadron-hadron
interactions.
The corresponding parton cascades are characterized
by a small momentum transfer and therefore, perturbative QCD cannot be
applied.
Consequently, to describe soft processes an object called phenomenological
soft Pomeron was introduced.
The amplitude for the Pomeron exchange  cannot be calculated from
first principles and, therefore, it is postulated and simply parameterized.

As the energy increases, the contribution of another type of processes called
semihard which are characterized by the appearance
of jets of hadrons with large $p_T$ becomes important.
In these processes some partons in the cascade appear with large momentum transfer
and the perturbative methods become applicable.
The concept of semihard Pomeron was proposed to describe
this mechanism \cite{Kal94,Dre99,Ost02}.
It includes the use of a soft Pomeron description for the
low virtuality part of the parton
cascade and the perturbative QCD techniques for the high virtuality part.

At present, there are two different approaches in use to describe high energy
hadronic collisions:
(1) the Gribov-Regge Theory (GRT), which employs the soft and semihard
Pomeron description \cite{Gri68},
and (2) the QCD eikonal (mini-jet) approach \cite{Gai85,Dur87}.

QGSJET01 is based on GRT and the quark-gluon string model.
The parton densities used in this model
are based on  pre-HERA data. The cross section for
diffraction dissociation is a constant fraction of the elastic
cross section. At present, this model is considered outdated.

At very high energies the number of partonic cascades becomes very
large and they start to overlap and interact between each other.
These nonlinear effects can be described by Pomeron-Pomeron coupling
in the low virtuality region.
They are taken into account in QGSJET-II,
which is a successor of QGSJET01. The parton densities in this
model updated after the analysis of Ref.\cite{Lun04} include the
correct behavior for the ratio of diffractive to elastic cross
sections, i.e. decreasing with increasing energy.

SIBYLL is based on the minijet approach to describe semihard processes.
The new version of SIBYLL, SIBYLL2.1 \cite{Eng99}, includes the GRT
and takes into account
the exchange of multiple soft Pomerons to describe
soft processes. To treat nonlinear effects this model assumes
that parton densities in the region of small virtualities
are completely saturated and that partons are produced for transverse momentum
larger than some cutoff that increases with energy.
Also the updated parton densities were implemented in the new version of
SIBYLL.

EPOS \cite{Wer06} is a recent implementation of GRT.
EPOS stands for Energy-conserving
quantum mechanical multiple scattering approach, based on Partons
(parton ladders), Off-shell remnants, and Splitting of parton
ladders. In this model, like in QGSJET, soft and semi-hard
Pomeron amplitudes are used. The nonlinear effects are taken into account
by an effective treatment of lowest order Pomeron-Pomeron interaction graphs.
This model describes very well detailed RHIC data and other
available data from high energy particle physics experiments. In
EPOS, energy conservation is considered in both cross section and
particle production calculations. An important  feature of this model
is the explicit treatment of the projectile and the target remnant hadronization
which leads to a more complete description of baryon and antibaryon production.

\section{Calculations}

The results presented in this paper can be understood as a sort of
a quantitative experiment as we analyze the statistical secondary
particle information produced by different hadronic packages with
the same input parameters and compare them with each other. The
input parameters include: (1) the type of primary particle;
(2) the type of target; (3) the energy of the primary particle,
$E_P$ and (4) the number of collisions $N_{coll}$.
Because the main component of the air is nitrogen, we choose this
nucleus as a representative target for our case of hadronic collisions that occur
within the Earth's atmosphere. Typical primary particles are nucleons (proton, neutron)
and charged pions; other hadronic projectiles are also possible but their number
is substantially smaller in the case of EAS.
The energies of the projectiles range from the minimum energy
supported by the corresponding models
(30 GeV for EPOS and QGSJET, and 100 GeV for SIBYLL),
up to the highest cosmic ray energies ($\approx 100$ EeV).

The number of collisions was determined taking into account (1) the run time
of each hadronic package, and (2) the necessity
to obtain good enough statistics for further analysis. The SIBYLL
package has a shortest running time and the number of collisions for
this package was taken to be 10 000. For QGSJET-II  we also analyzed
10 000 collisions. For EPOS, 3 000
collisions were analyzed for lower energies, and for high
energies, when the number of secondary particles gets very large
and therefore the calculation gets very slow, $N_{coll}=1000$ was taken.

Each secondary particle is characterized by the
following: (1) its type, e.g. proton, $\pi^+$, etc.; (2) its kinetic
energy $E_{sec}$; (3) the angle between the primary and the secondary
particle direction. All the observables discussed in this work correspond to the
laboratory system.

It is known that the secondary particles with small energies
(say, less than 40 MeV) do not contribute significantly to the air 
shower development
and are not tracked in EAS simulations; therefore, such particles
are excluded from the present analysis.

We separated all collision events into inelastic and those what
we call VELP events.
As it was already mentioned in the introduction, in this work we
are interested in the processes which are effective as a way to transport
the energy deep down the atmosphere. To separate the VELP
events we apply the following criterion:
(1) among the secondaries produced after a given collision 
with primary energy $E_P$,
the most energetic one is localized and labeled
as the ``leader'' (or leading particle), carrying the energy $E_{lead}$.
(2) The average energy of the rest of the secondaries
(those not including the leading particle), $\langle E_{sec}\rangle$, is determined.
(3) The leading energy fraction, defined as
\begin{equation}
f_L=\frac{E_{lead}}{E_P},
\end{equation}
is then analyzed as follows:
\begin{itemize}
\item [(i)] if $f_L\geq f_1$ the event is labeled as a VELP one ($f_1$ is a given constant).
\item [(ii)] if $f_2<f_L<f_1$, ($f_2$ another given constant)
the event is labeled as a VELP one only if
$\langle E_{sec}\rangle$/$E_{lead}$ is larger than a given value $g$ that depends on $f_L$.
\item [(iii)] In any other case the event is labeled as non VELP or
``inelastic collision''.
\end{itemize}

In all our calculations we have taken $f_1$ = 0.95 and  $f_2$ = 0.3,
and $g(f_L)=0.01+0.6(1-f_L)^2$. This particular election corresponds
to an efficient way of labeling events characterized by a relatively
small number of secondaries containing an energetic leading particle
 capable of contributing considerably to the energy transport deep
down in the atmosphere during the air shower development.
The VELP events distinguished with the above criterion
include most of the standard diffraction events.

\begin{figure}[p]
\centerline{\includegraphics{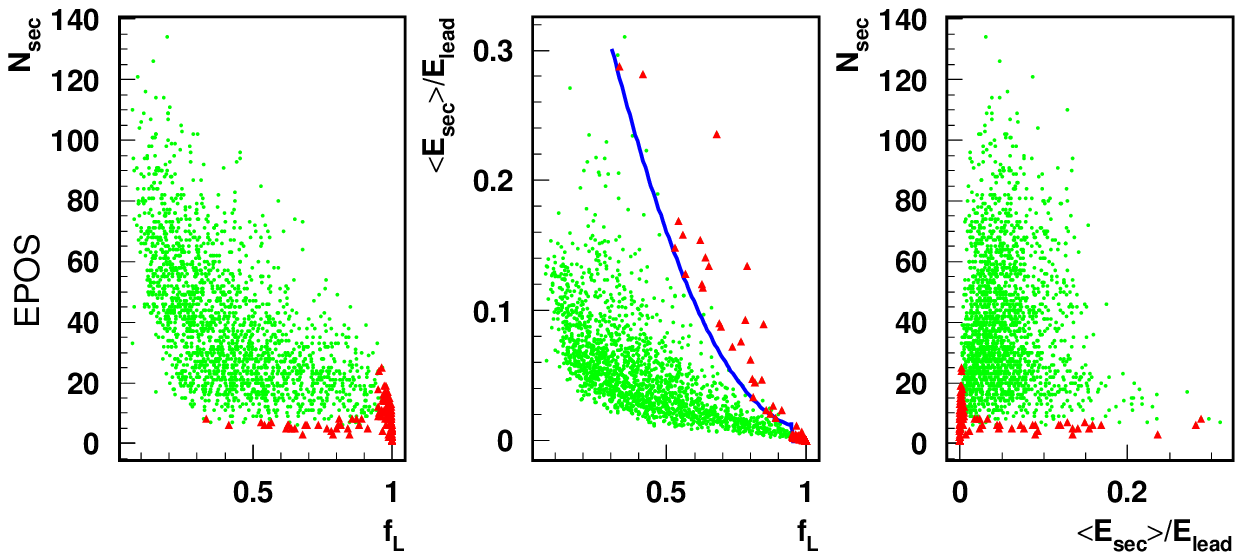}}
\centerline{\includegraphics{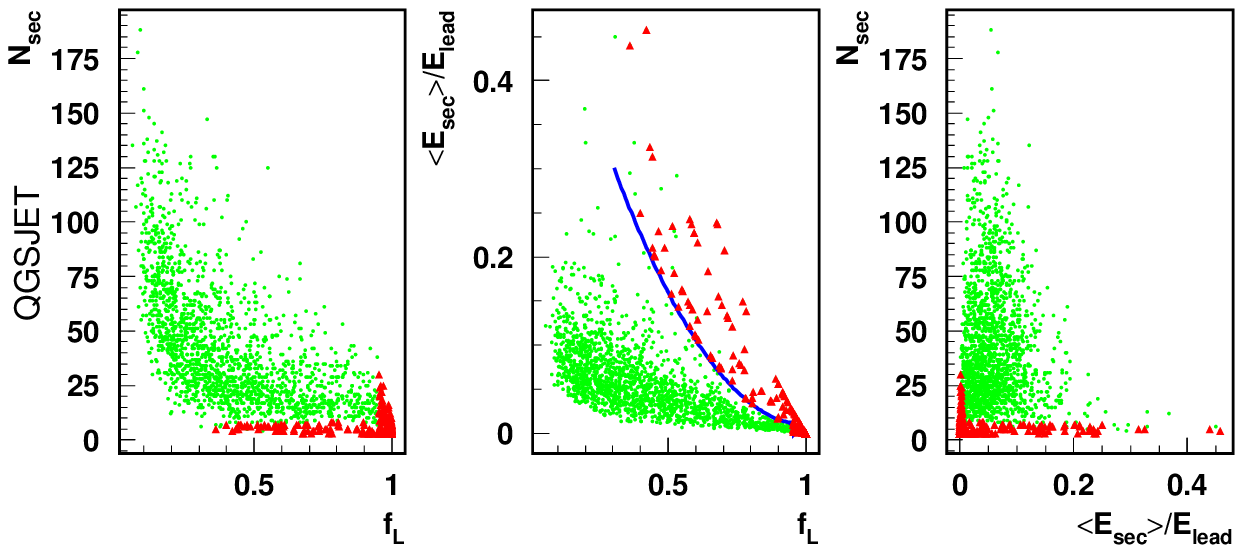}}
\centerline{\includegraphics{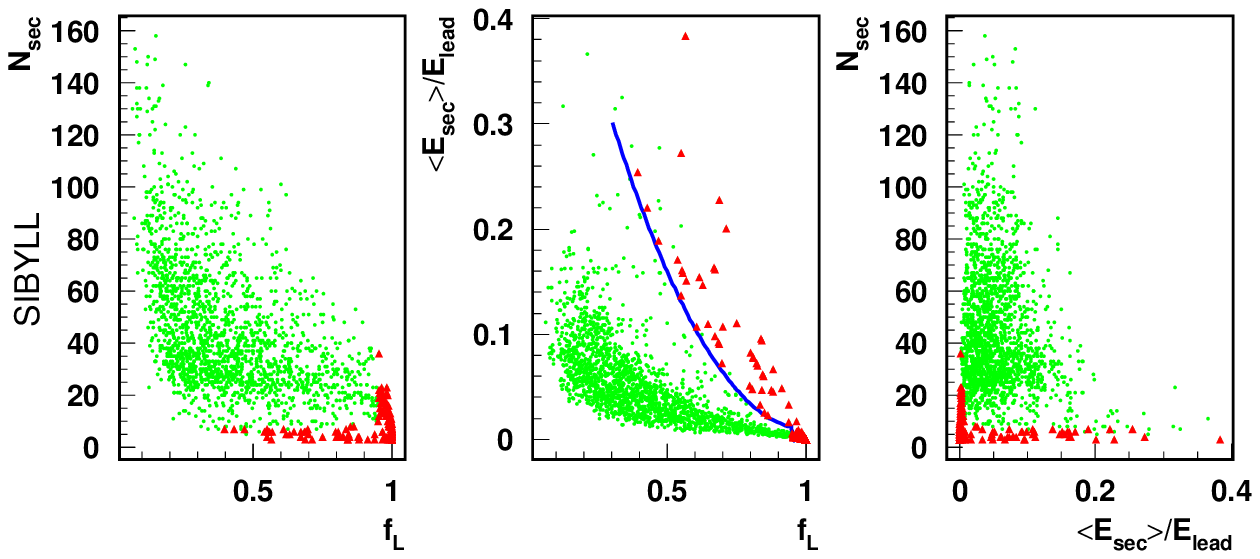}}
\caption{(color online). Scatter plots illustrating VELP event
  selection, produced with sets of 2000 collisions of 10 TeV protons,
  simulated with EPOS (upper row), QGSJET-II (middle row), and SIBYLL
  (lower row). The plots in the left side column correspond to
  $N_{sec}$ versus $f_L$, while the middle and right side columns
  correspond to $\langle E_{sec}\rangle$/$E_{lead}$ versus $f_L$ and $N_{sec}$
  versus $\langle E_{sec}\rangle$/$E_{lead}$, respectively. The large red
  triangles (small green dots) correspond to VELP (non-VELP) events;
  the solid line in the middle column plots represents the function
  $g(f_L)$ (see text).}
\label{criterion}
\end{figure}

To illustrate how our algorithm works we present in Fig.\ref{criterion} 
a representative case that corresponds to
proton-nitrogen collisions at energy $E_P=10$ TeV calculated using
EPOS (upper row), QGSJET-II (middle row) and  SIBYLL (lower row).
The VELP events are shown by large red triangles whereas the rest
of the events correspond to small green dots. The plots in the
left side column represent the $N_{sec}$ versus $f_L$ distributions
of the collision events. As it was already mentioned before, all
events with $f_L\geq 0.95$ are marked as VELP. They lie approximately along
a straight line and are characterized by a relatively
low multiplicity ($N_{sec}\lesssim 20$). When $f_L$ is not that
large ($0.3 < f_L < 0.95$) there is a number of events that are
characterized by a very low multiplicity ($N_{sec} < 10$) and therefore
should be chosen as VELP. These events lie on the approximately straight
line parallel to the $f_L$-axis of the plot. It is important to notice that
in the case of EPOS there are very few events with $N_{sec} < 10$
and $0.3 < f_L < 0.95$ compared to case of QGSJET-II and SIBYLL.

On the plots in the middle of Fig.\ref{criterion} the $\langle
E_{sec} \rangle/E_{lead}$ versus  $f_L$ distributions are shown. The
events with $f_L\geq 0.95$ are characterized by a small value of
$\langle E_{sec}\rangle$/$E_{lead}$ where they concentrate. 
The VELP events corresponding to $0.3 < f_L < 0.95$
lie above the solid line representing the function $g(f_L)$ defined above.
The form of this function was obtained empirically so that the 
events with low multiplicity and large $f_L$ are VELP's  in
the simulations with all packages: EPOS, QGSJET-II and SIBYLL.
Notice that above the solid line there are non-VELP events.
These are the events where the leading particle possess the quantum numbers
different from that of the projectile and therefore they are marked as 
``inelastic''.

The plots on the right side show the $N_{sec}$ versus $\langle E_{sec}
\rangle/E_{lead}$ distributions. It is seen that large values
of $\langle E_{sec}\rangle$/$E_{lead}$ correspond to  events
with low multiplicity.

%
%



\begin{figure}
\centerline{\includegraphics[scale=0.7]{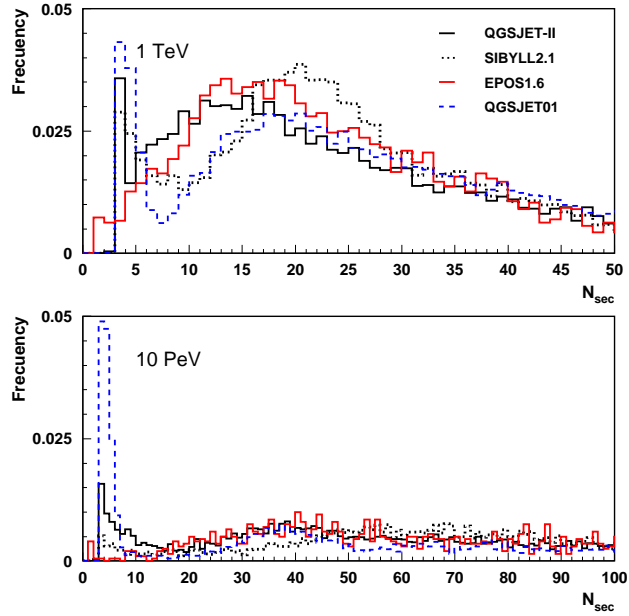}}
\caption{(color online).  Number of secondary
particles distributions at two different energies: 1 TeV and 10
PeV for the proton-nitrogen collisions.}
\label{nsec}
\end{figure}

%
%
We start our analysis from the studies of the multiplicity
 of secondary particles.
In Fig.\ref{nsec} we show the distributions of the number of
secondary particles $N_{sec}$ corresponding to events at two
energies. The plotted frequency is the number of events with the
specific $N_{sec}$ divided by the total number of collisions. One
can identify the peak at low $N_{sec}$ as the signature of
VELP events. It can be seen that for small $N_{sec}$  the
shape of the distribution is different for all models.
It is worthwhile mentioning that the data generated with EPOS1.6
does not present a prominent VELP peak for low multiplicity, compared
with the other models. At high energies, the distribution gets flatter
and the VELP peak is clearly seen, but it is small.
From the EPOS1.6 documentation that is available, we cannot find a
clear explanation for this different behavior. Nevertheless,
we consider that this feature of EPOS could come from the
generation of diffractive events that contain a not very reduced number
of secondaries. 

%
%

\begin{figure}
\centerline{\includegraphics[scale=0.7]{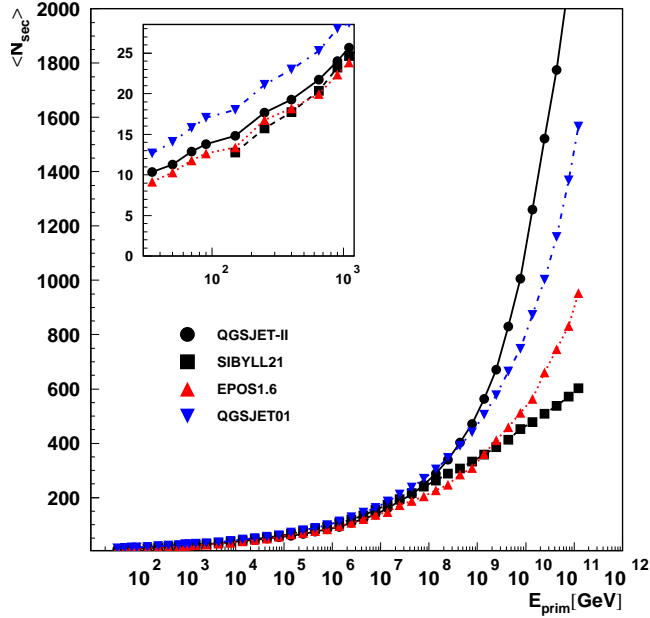}}
\caption{(color online). The dependence of the
mean number of secondary particles ($<N_{sec}>$) on the primary
energy for the case of proton-nitrogen collisions.} \label{avnsec}
\end{figure}

It is also interesting to compare the energy dependence of
averaged multiplicities calculated in different models. In Fig.\ref{avnsec} 
we show the average number of secondary particles
$\langle N_{sec}\rangle$ produced in the collisions as a function of the primary
energy. It can be seen that for $E_P>$10$^8$ GeV the
difference between the two models becomes significant. At highest
energies the largest amount of secondary particles is produced in
the QGSJET-II case, clearly larger than the cases of QGSJET01c and EPOS1.6.
The least $\langle N_{sec}\rangle$ is given by SIBYLL2.1. For example, at 10 EeV
SIBYLL2.1 produces in average approximately 425 secondary
particles, EPOS1.6 -- 450, QGSJET01c -- 675 particles and
QGSJETII-3 produces 1225 particles. This high number of
secondary particles produced at the highest primary energies is a
well known feature of the QGSJET package.

For low energies the largest amount of secondaries is given by
QGSJET01c, and then by QGSJETII-3, EPOS1.6, SIBYLL2.1. EPOS1.6
produces more secondary particles than QGSJET-II. But the
difference is not as large as in the high energy case.

%
%

\begin{figure}
\centerline{\includegraphics[scale=0.7]{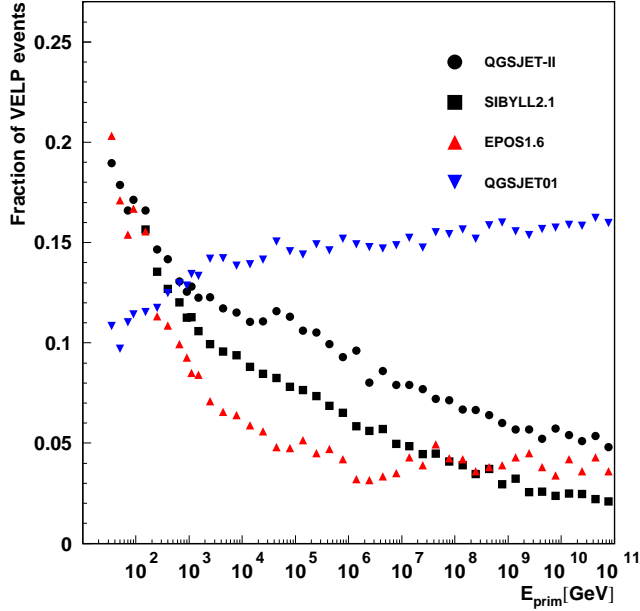}}
\caption{(color online).Fraction of VELP events versus
primary energy for the case of proton-nitrogen collisions.}
\label{diffrac}
\end{figure}

In Fig.\ref{diffrac} we present the dependence of the fraction of
VELP events on the primary energy. The fraction of
VELP events is defined as the number of VELP events
divided by the total number of events and this quantity is directly
related to the diffractive to total cross section ratio.

All the models studied, except QGSJET01c, present a similar shape
for the fraction of VELP events: it reduces with the
primary energy. On the contrary, as it was already discussed in
Ref.\cite{Lun04}, QGSJET01c shows a nearly constant dependence.
The QGSJETII-3 gives a larger amount of VELP events than
SIBYLL2.1 and EPOS1.6. This is expected from the pronounced
VELP peak in the multiplicity distributions generated by
QGSJETII-3. As before, for the energy of 1 TeV QGSJET and
SIBYLL2.1 give similar results. At the highest energies QGSJETII-3
gives the highest fraction (5.5$\%$), SIBYLL2.1 (3$\%$) the least,
EPOS1.6 is in between (4$\%$). EPOS1.6 produces less VELP
events for lower energies and shows weaker dependence on energy
for high primary energies compared to other models. Indeed, for
very high energies the dependence is almost flat.

For very high energies, the information on diffractive to total
cross sections ratios obtained from the shower development is of
important interest for particle physics because accelerator data is
unavailable for this energy range. There are different theoretical
models that predict diffractive cross sections and they all differ
substantially for very high energies (see, for example,
Refs.\cite{Got95, Kay04,Tro05} and references therein). Therefore,
the cosmic ray data could help to distinguish among these models.

%
%

\begin{figure*}
\centerline{\includegraphics[scale=1.2]{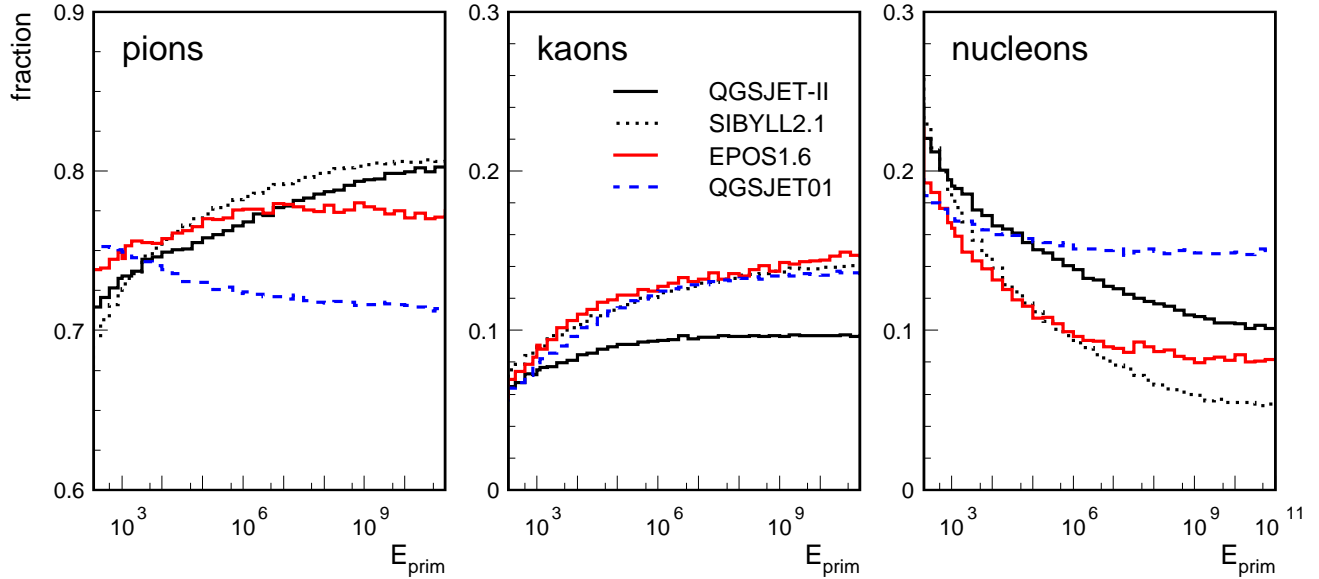}}
\caption{(color online).The dependence of the
fraction of secondary pions, kaons and nucleons on the primary
energy in the case of proton-nitrogen collisions.} \label{mfrac}
\end{figure*}

We now turn to the study of the relative amount of
different secondary particles produced in the collisions. On
Fig.\ref{mfrac} we present the dependence of the fraction of
secondary pions, kaons and nucleons on the primary energy. This
fraction is defined as the mean number of secondary particles
of a given type $\langle N_{sec}^i\rangle$ (where $i$ stands for the
particle type) divided by the mean total number of
secondary particles $\langle N_{sec}\rangle$.

It can be seen that at the highest energies approximately 80$\%$ of all
secondary hadrons are pions (neutral and charged). Pions are
important for shower development because: (1) charged pions decay
into charged muons which are detected by surface detectors;
(2) neutral pions decay into gamma quanta and thus initiate
electromagnetic showers.
It can be seen that SIBYLL2.1, QGSJETII-3 and EPOS1.6
produce similar results, that is a pion fraction that increases with energy
and saturates at the highest primary energies. At $10^{11}$ GeV SIBYLL2.1 and
QGSJETII-3 produce virtually identical results,
that are 7$\%$ larger than the EPOS1.6 fraction.
On the other hand, QGSJET01 predicts a fraction of pions that
decreases with energy.

Kaons also play a significant role in shower development (neutral kaons decay
into neutral pions and charged kaons decay into charged pions,
which in turn decay into detectable muons)
and it is thus worthwhile studying their production rates.
All models produce kaon fractions slowly
increasing with energy. At the highest energies, kaons represent
approximately 14$\%$ (QGSJETII-3 gives 10$\%$) of all secondaries.

The right hand panel of Fig.\ref{mfrac} shows the dependence of
the secondary nucleon and antinucleon fraction
on the primary energy. It can be seen that the fraction of such particles
reduces with primary energy.  In our analysis we see that the
largest amount of baryons is produced by the QGSJETII-3
package (10$\%$), the least is given by SIBYLL2.1 (5$\%$) and
EPOS1.6 (8$\%$) is in between of both. It can be seen that QGSJETII-3
produces more nucleons than SIBYLL2.1 and EPOS1.6.

%
%

\begin{figure*}
\centerline{\includegraphics[scale=1.2]{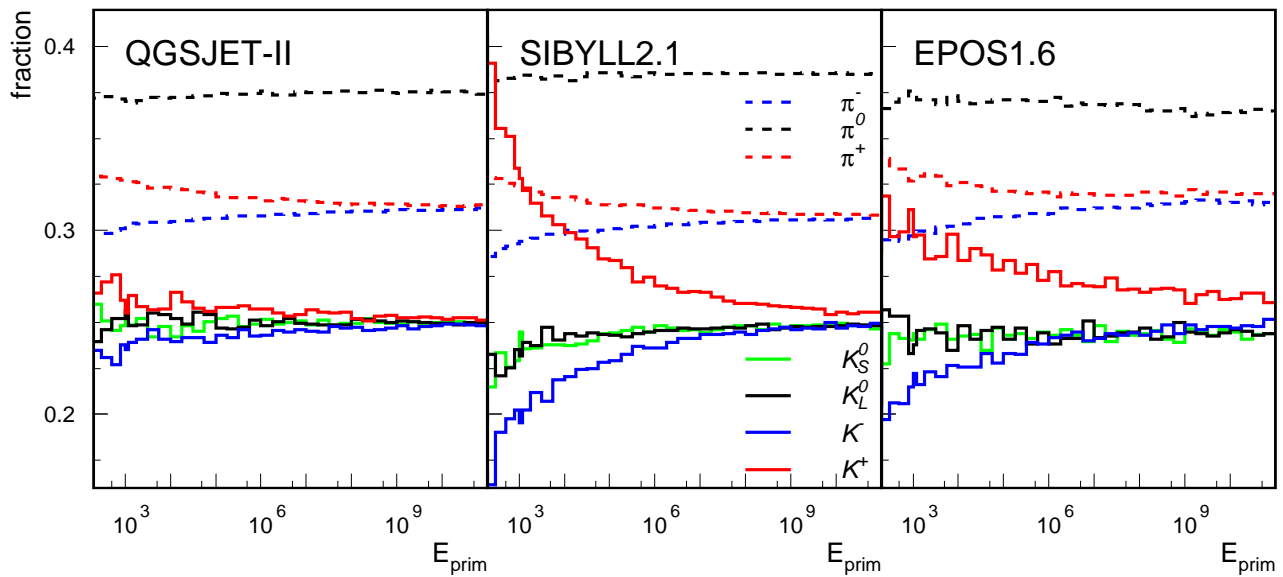}}
\caption{(color online). The dependence of the
fraction of secondary pions and kaons separated by charge on the
primary energy in the case of proton-nitrogen collisions. } \label{mmfrac}
\end{figure*}

Fig.\ref{mmfrac} shows the relative amount of pions and kaons
separated by charge for QGSJETII-3, SIBYLL2.1 and EPOS1.6. We show
the energy dependence of the relative fractions of $\pi^-$,
$\pi^0$, $\pi^+$ and $K^-$, $K^0_S$, $K^0_L$, $K^+$.

Notice that the fraction of pions of each type is defined as its mean
multiplicity divided by the total mean multiplicity of pions; and similarly
the $K^-$, $K^0_S$, $K^0_L$, $K^+$ fractions add up to 1.

It is seen that 38$\%$ of all pions are neutral pions. The
relative amount of neutral kaons does not change with primary
energy. At high energies the amount of $\pi^-$ and $\pi^+$ becomes
equal (31$\%$ of all pions). At lower energies there are more
$\pi^+$ than $\pi^-$.
There are no important differences between different models with
respect of the multiplicity of different charges of pions.

In the case of kaons, it is found that
at high energies each type of kaons contribute 25 $\%$
to total kaon multiplicity. The fractions of $K^0_S$ and $K^0_L$
do not change significantly with energy. At lower primary energies, there are
more $K^+$ than $K^-$. The models differ with respect to the slope
of energy dependence of the fractions of $K^-$ and $K^+$.

%
%

\begin{figure*}
\centerline{\includegraphics[scale=1.2]{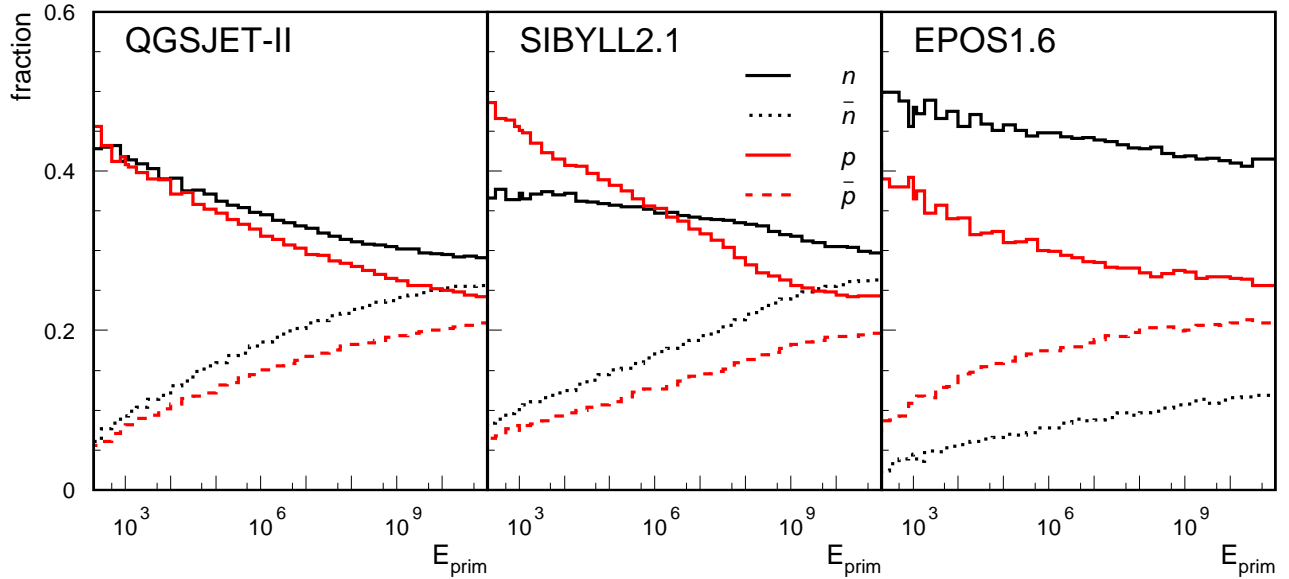}}
\caption{(color online). The dependence of the
fraction of secondary nucleons separated by charge on the primary
energy in the case of proton-nitrogen collisions.} \label{mbfrac}
\end{figure*}

On Fig.\ref{mbfrac} we show the relative fractions of
secondary neutrons, antineutrons, protons and antiprotons as a
function of the primary energy. These fractions are defined as the
mean multiplicity of the particle of each charge divided by the
total mean multiplicity of nucleons and antinucleons.
We recall that in the analysis of these secondaries, we have excluded all
particles with kinetic energy less than 40 MeV, due to their
irrelevance in the case of air shower development.
Such low energy particles are, in general, nucleons.

It can be seen that the amounts of neutrons and protons decrease
with energy, whereas the amounts of antineutrons and antiprotons increase.
Notice that at the highest energies both QGSJETII and SIBYLL2.1 produce
similar amounts, i.e. 25$\%$ approximately, of $n$, $\overline n$, $p$,
$\overline p$; this is not the case of EPOS1.6 (see below).

The slopes of the plots produced in three models are different
from each other. EPOS1.6 shows the largest separation between the
fractions of nucleons and antinucleons at high energies and for
all energies EPOS produces more neutrons and less antineutrons.
Namely, at high energies EPOS1.6 produces approximately 30$\%$
more neutrons and 70$\%$ less antineutrons than, for example,
SIBYLL2.1. For secondary protons and antiprotons the situation is
different. For low energies, EPOS1.6 produces less protons
compared to SIBYLL2.1 and QGSJETII-3. At high energies ($E_p>500$
PeV), EPOS1.6 gives slightly more protons than SIBYLL2.1 and
QGSJETII-3.
Notice also that in the QGSJETII-3 and SIBYLL cases and for all energies
the fraction of antineutrons is larger than the corresponding one for antiprotons.
This contrasts with the EPOS1.6 case where the fraction of antiprotons
is larger than the fraction of antineutrons.

%
%

Now we turn our attention to the energy of secondary particles
produced in proton-nitrogen collisions.
We study the fraction of interaction energy carried by secondary particles.
This fraction is defined as the mean energy carried by secondary particles of a
certain type divided by the primary energy.

\begin{figure}
\centerline{\includegraphics[scale=1.3]{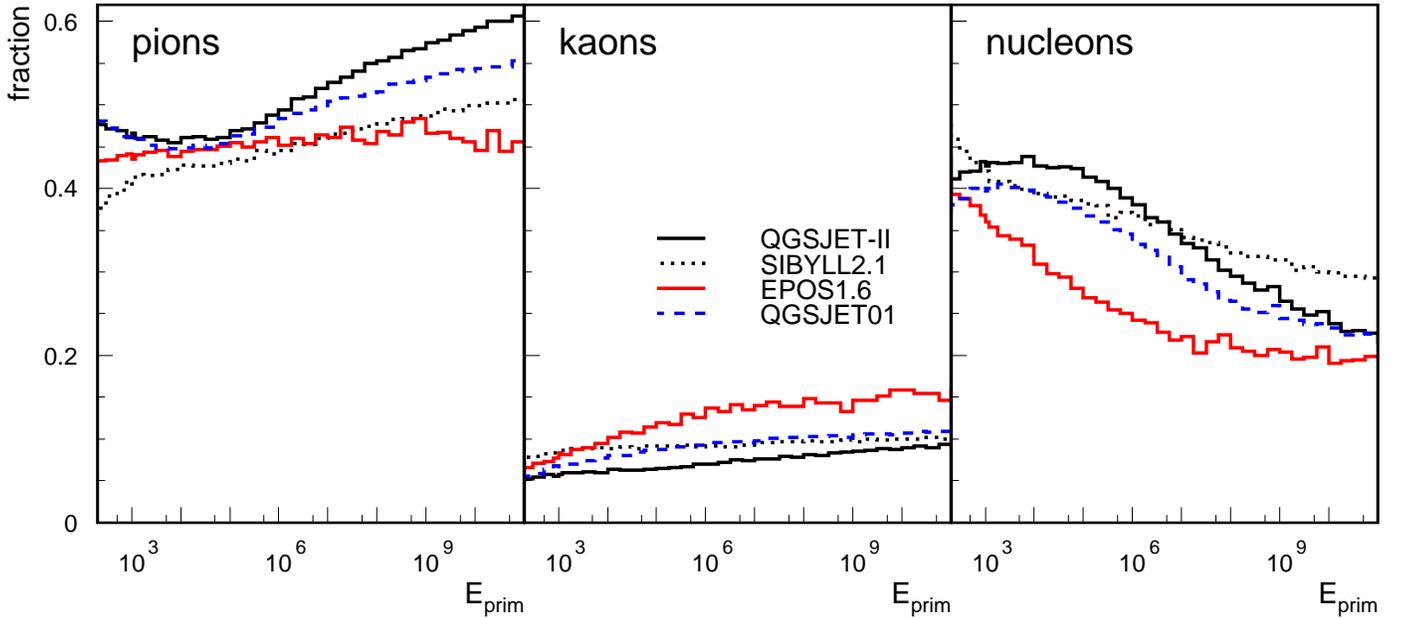}}
\caption{(color online). Fractions of
interaction energy carried by secondary pions, kaons and nucleons
as a function of the primary energy in the case of proton-nitrogen
collisions. }
\label{e-per3-test}
\end{figure}

In Fig.\ref{e-per3-test} we compare the fraction of mean secondary energy
carried by pions, kaons and nucleons.
Generally these plots follow the behavior of the mean multiplicity
of pions, kaons and nucleons presented in Fig.\ref{mfrac}. That is,
the relative amount of energy and the general shape of the primary energy
dependence are dictated by multiplicity plots.

The mean multiplicity of pions increases with energy and consequently
increases the amount of energy carried by pions and the same is true for kaons.
The mean multiplicity of nucleons decreases with primary energy and the fraction of
interaction energy decreases as well.

Notice that the major part of the interaction energy is
carried by pions which is expected from the fact that the majority
of secondary particles are pions. Pions produced in QGSJETII-3
carry the largest amount of energy compared to other models. At
largest energies pions carry approximately 60$\%$ (QGSJET23),
55$\%$ (QGSJET01c), 50$\%$ (SIBYLL2.1), 45$\%$ (EPOS1.6).

In EPOS case, kaons take away approximately 5$\%$ more energy than
in other models. For example, at the largest energies, the  kaons produced by EPOS1.6
 carry approximately 15$\%$ of the primary energy, in contrast with 10$\%$ for QGSJETII-3 or
SIBYLL2.1 or QGSJET01c.

At highest energies nucleons produced in SIBYLL2.1 carry
approximately 30$\%$ of energy, QGSJETII-3 and QGSJET01c - 23$\%$
and in EPOS1.6 20$\%$. For all primary energies the amount of
energy carried by nucleons produced in EPOS1.6 is the smallest.

%
%

On the next two figures we present the fractions of interaction energy
carried by secondary mesons and nucleons separated by charge.

\begin{figure}
\centerline{\includegraphics[scale=1.3]{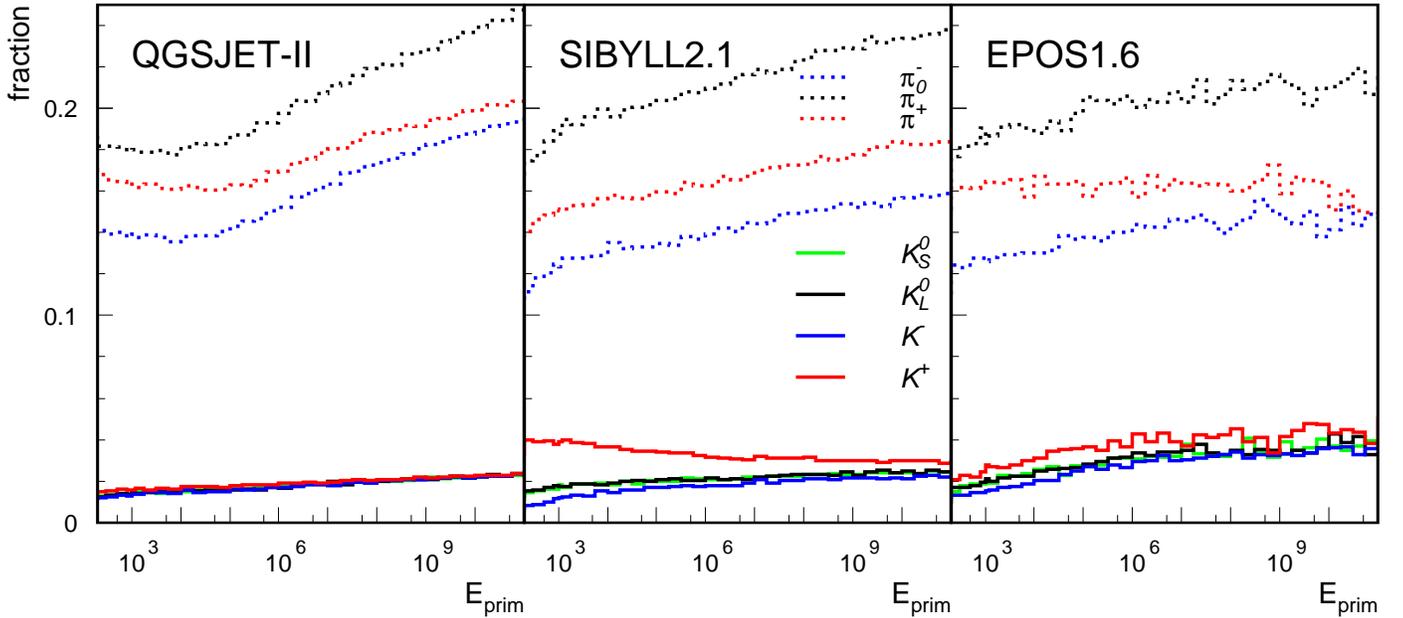}}
\caption{(color online). Fractions of
interaction energy carried by different secondary mesons as a
function of the primary energy in the case of proton-nitrogen
collisions.}
\label{e1-mesons-test}
\end{figure}

In Fig.\ref{e1-mesons-test} we show the fraction of mean secondary
energy carried by mesons: pions and kaons generated in
QGSJETII-3 (left panel), SIBYLL2.1 (middle panel) and EPOS1.6 (right
panel).
 As before, the fraction of interaction
energy carried by the particles of each type is defined as the
total energy carried by these particles divided by the primary
energy.

As expected from the multiplicity plots, the largest fraction
of energy is carried by neutral pions and its value
increases with primary energy.
The curves for charged pions calculated using
EPOS160 present almost flat dependence.

In the case of kaons, all kaons show the same energy dependence
slowly increasing with energy. Only the fraction of interaction
energy carried by K$^+$ calculated by SIBYLL2.1 decreases with
energy. This can be explained by the large multiplicity of
positively charged kaons generated in this package.

%
%

\begin{figure}
\centerline{\includegraphics[scale=1.3]{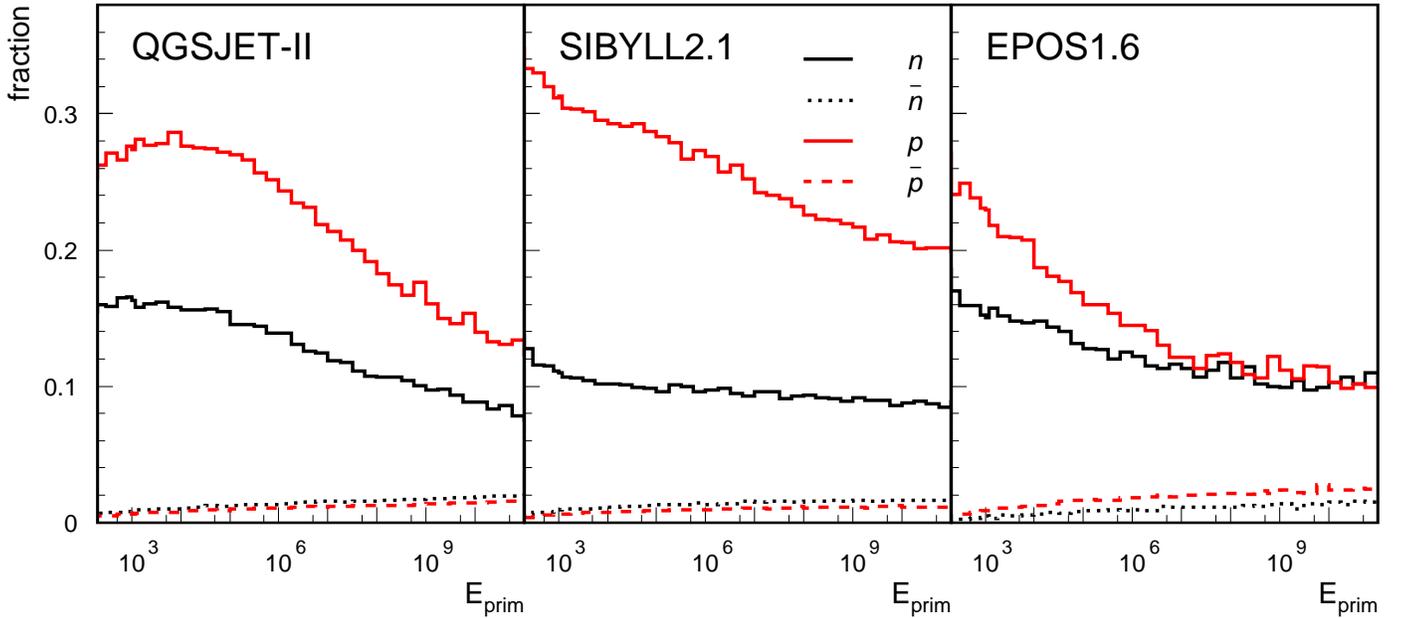}}
\caption{(color online). Fractions of interaction energy
carried by secondary nucleons as a function of the primary energy
in the case of proton-nitrogen collisions. }
\label{e1-nucleons-test}
\end{figure}

In Fig.\ref{e1-nucleons-test} we show the fraction of mean
secondary energy carried by neutrons, antineutrons, protons and
antiprotons generated in QGSJETII-3 (left panel), SIBYLL2.1 (middle
panel) and EPOS1.6 (right panel).

In all models, the largest amount of energy is carried by protons.
This is expected from the fact that proton is a projectile.
Neutrons carry smaller fraction of interaction energy and antiprotons and antineutrons
carry very small fraction of interaction energy.

The fraction of interaction energy decreases with primary energy for protons and neutrons
and slowly increases for antiprotons and antineutrons. This is in accordance with the
multiplicity plots for baryons.

It is seen that in the case of EPOS1.6, protons and neutrons carry
the same fraction of interaction energy for $E_{prim}> 10$ PeV.
Generally, all models give similar values of fraction of
interaction energy carried by neutrons. For protons, there is a
difference between models. For all energies, the protons generated
in EPOS carry less interaction energy. This can be explained by
less amount of VELP events seen in EPOS1.6.

%
%

\begin{figure}
\centerline{\includegraphics[scale=0.7]{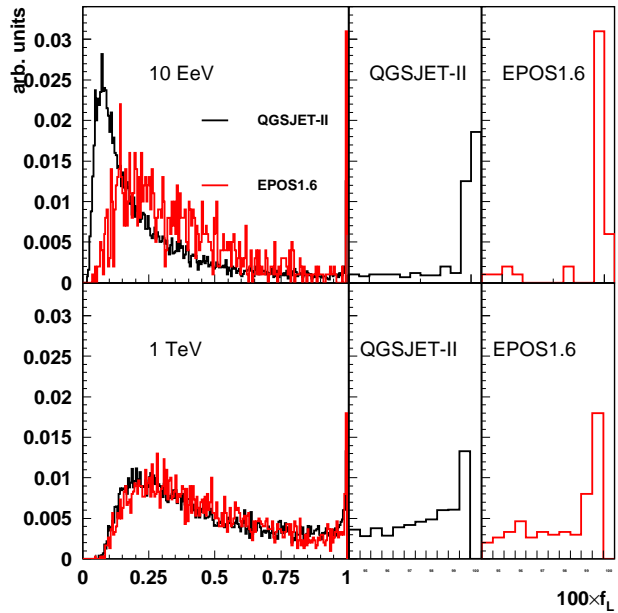}}
\caption{(color online). Leading energy
fraction distributions for proton-nitrogen collisions at
$E_{prim}$ = 10 EeV(upper panel) and $E_{prim}$ = 1 TeV(lower
panel). }
\label{lead}
\end{figure}

In Fig.\ref{lead} we show the distribution of  the leading energy
fraction $f_L$ for two values of the primary energy, namely 1 TeV
and 10 EeV. $f_L$ distributions are important when one wants to
separate the VELP events from the ``inelastic'' collisions. For
clarity, we compare only the results of QGSJETII-3 and EPOS1.6.
There are two peaks in this distribution. Most of the VELP events
come from  the relatively sharp peak at $f_L= 1$, which
corresponds to the existence of a leading particle among few
secondaries which carries almost all available energy. The second
peak is wide and its position changes with energy. As the primary
energy increases, the number of secondary particles grows and the
wide maximum of the distribution shifts towards $f_L=0$. It is
seen from this plot that at 10 EeV, QGSJETII-3 has a maximum very
close to $f_L=0$, which can be explained by the very large number
of relatively low-energy secondaries generated in this package. At
low energies, the VELP peak is less pronounced. As the energy goes
up, the peak generated by all models becomes more pronounced, and
at highest energies EPOS1.6 produces the tallest peak
corresponding to VELP events. 

%
%

We conclude this section with the study of the pseudorapidity
distributions of secondary particles. They are important in cosmic
ray physics because they are significantly correlated with the
lateral distributions of muons at large distances from the core.
Pseudorapidity is defined by $\eta = -\ln\tan\frac{\Theta}{2}$,
where $\Theta$ is the angle that specifies the direction of motion
of a secondary particle with respect to the direction of the
primary particle. Using the pseudorapidity allows to distinguish
the secondary particles by their direction of motion with respect
to the primary particle, which is important for air shower
development.

\begin{figure*}
\centerline{\includegraphics[scale=1.2]{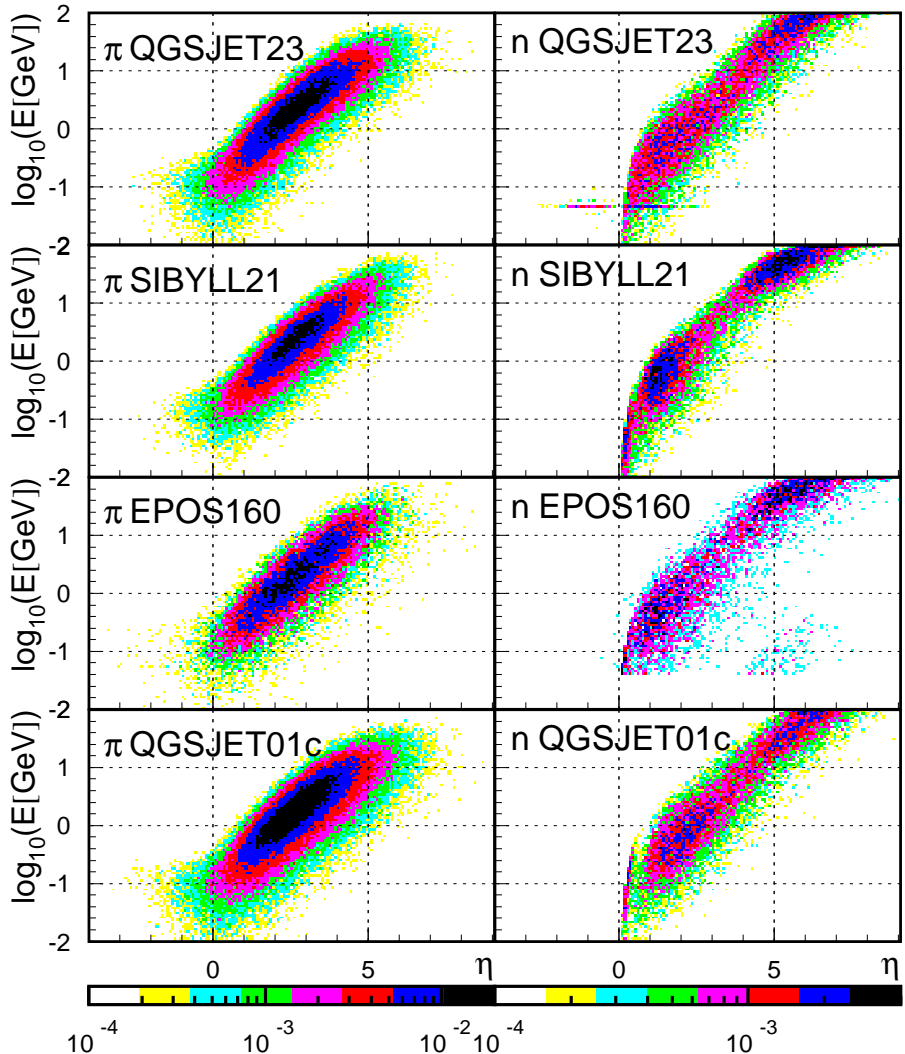}}
\caption{(color online).  Pseudorapidity of secondary pions(left
column) and nucleons(right column) versus their energy for the
case of proton-nitrogen collisions at 1 TeV. Note different scale
for pions and nucleons.} \label{pseudorap1}
\end{figure*}

In Fig. \ref{pseudorap1} we present the $\eta\times E_{sec}$
two-dimensional distributions for proton-nitrogen collisions at 1
TeV. The left panels show the distributions of the secondary pions
and the right panels those of secondary nucleons.

The most outstanding characteristic of the plots in Fig. \ref{pseudorap1}
 is the clear linear behavior of the mean pseudorapidities at a given energy
with the logarithm of the secondary energy. The slopes of the corresponding
lines are similar for all models.

In the case of pions (left hand side plots) the distributions
possess a simple structure. The pseudorapidity distributions
at a fixed secondary energy are approximately
gaussians with energy independent standard deviation.

The distributions for nucleons present two peaks, consequence of the
bimodal energy spectra of secondary nucleons, that is characteristic
of all hadronic models \cite{Lun04}.  In all models the pseudorapidity
distributions in the zones where $\eta$ is around zero, or negative
(recoiling particles) are somewhat unnatural (see right hand column of
Fig. \ref{pseudorap1}), presenting a relative abundance of particles
with positive but very small $\eta$, and zero recoiling particles
($\eta < 0$) \cite{foot1}. EPOS1.6 distributions for nucleons also
indicate the existence of low energy secondaries in the near forward
direction ($\eta > 4$).  From the available EPOS1.6 documentation we
cannot find any explanation for such particles.

\section{Conclusions and outlook}
We have performed a comparative analysis of secondary particle
observables produced in collisions generated by different hadronic
packages. We studied the secondary particle information generated
by SIBYLL2.1, QGSJETII-3 and EPOS1.6 using identical input data:
projectile and target type and primary energy. For comparison with
previous analysis, we also included the results of QGSJET01c, even
if this package is already considered outdated.

The choice of studied quantities was dictated by
their importance for the air shower development.
We studied multiplicity distributions of secondary particles, mean multiplicity,
inelasticity, fraction of secondary pions and baryons,
energy distributions of secondary particles and pseudorapidity distributions.
It was shown that the studied models present significant differences for all energy
ranges.

We introduced the notion of Very Energetic Leading secondary Particles (VELP) events,
corresponding to events that are characterized by a small
number of secondaries and that include a leading particle that carries
a substantial fraction of the projectile energy.
These events in their majority correspond to the standard diffractive events.

It is seen that QGSJETII-3 produces a very large amount of
secondary particles at high energies, twice as much as EPOS1.6.
Also, the multiplicity distributions show a different shape for
the VELP peak in the case of EPOS1.6 where the peak is
small compared to the results generated by other packages.

Because of their importance for shower development, a special
attention was given to VELP  processes. Our results for the
fraction of VELP events at the highest energy are:
QGSJETII-3 gives the highest fraction (5.5$\%$), SIBYLL2.1 (3$\%$)
the least, EPOS1.6 is in between (4$\%$). Our analysis gives larger
values for the fraction of VELP events than those obtained
in Ref.\cite{Lun04}. The primary energy dependence of the fraction
of VELP events becomes almost flat at high energies. This
implies the almost constant diffractive to total cross section
ratio for high energies. For lower energies EPOS generates less
VELP collisions when compared to the  other models.

Our analysis has shown that EPOS1.6 generates less pions than the
other packages (7$\%$ less than SIBYLL2.1 at 10 EeV). Another
feature of EPOS1.6 is the fraction of secondary neutrons and
antineutrons: our calculation gives 30$\%$ more neutrons and
70$\%$ less antineutrons at $E_P = 10^5$ GeV.

The next step in this analysis is the study of the impact of
different models of hadronic interactions on common air shower
observables \cite{Tar08}. To simulate air shower development the
AIRES program \cite{Sci01} is being used with QGSJETII-3 and
EPOS1.6 packages for the generation of hadronic interactions.

\acknowledgements
This work was partially supported by ANPCyT and CONICET, Argentina.


\begin{thebibliography}{99}
\bibitem{Kna03} J. Knapp, D. Heck, S.J. Sciutto, M.T. Dova, and M. Risse, Astropart. Phys.
{\bf 19}, 77 (2003).
\bibitem{Fle94} R.S. Fletcher {\it et al.}, Phys. Rev. D {\bf 50} 5710, (1994).
\bibitem{Eng99} R. Engel, T.K. Gaisser, T.Stanev and P. Lipari, in
{\it  Proc. 26th Int. Cosmic Ray Conf.}, Salt Lake City (USA),{\bf 1}, 415, (1999).
\bibitem{Kal97} N.N. Kalmykov {\it et al.},
Nucl. Phys. B (Proc. Suppl.) {\bf 52} 17, (1997)
\bibitem{Ost06} S. Ostapchenko, Phys. Rev. D {\bf 74} 014026, (2006)
\bibitem{Wer06} K. Werner, F.M. Liu and T. Pierog,
Phys. Rev. C {\bf 74} 044902, (2006); [arXiv:hep-ph/0506232].
\bibitem{Dre01} H.J. Drescher, M. Hladik, S. Ostapchenko,
T. Pierog and K. Werner , Phys. Rep. {\bf 350} 93, (2001)
\bibitem{Bar02} V. Barone, E. Predazzi, {\it High-Energy Particle Diffraction},
 Springer, 2002
\bibitem{Lun04} R. Luna, A. Zepeda, C.A. Garc\'ia Canal and S.J. Sciutto,
Phys. Rev. D {\bf 70} 114034, (2004).
\bibitem{Kal94} N.N. Kalmykov, S.S. Ostapchenko and A.I. Pavlov:
Bull. Russ. Acad. Sci. Phys. {\bf 58} 1966, (1994); Nucl. Phys. Proc. Suppl. B {\bf 52}
17, (1997).
\bibitem{Dre99} H.J. Drescher {\it et al.}, J.Phys. G {\bf 25} L91, (1999).
\bibitem{Ost02} S. Ostapchenko {\it et al.}, J.Phys. G {\bf 28} 2597, (2002).
\bibitem{Gri68} V.N. Gribov, Sov. Phys. - JETP {\bf 26} 414, (1968);{\bf 29} 483, (1969).
\bibitem{Gai85} T.K. Gaisser and F. Halzen, Phys. Rev. Lett.{\bf 54} 1754, (1985);
G. Pancheri and Y. Srivastava, Phys. Lett. {\bf 159B} 69, (1985).
\bibitem{Dur87} L. Durand and H. Pi, Phys. Rev. Lett. {\bf 58} 303, (1987);
Phys. Rev. D {\bf 38} 78, (1988).
\bibitem{Got95} E. Gotsman, E.M. Levin, and U. Maor,
Phys. Lett. B {\bf 347}, 424, (1995);  [arXiv:hep-ph/9407227].
\bibitem{Kay04} A.B. Kaydalov, V.A. Khoze, A.D. Martin, and M.G. Ryskin,
Eur. Phys. J. C {\bf 33}, 261, (2004);  [arXiv:hep-ph/0311023].
\bibitem{Tro05} S.M. Troshin and N.E. Tyurin,
Eur. Phys. J. C {\bf 39}, 435, (2005);  [arXiv:hep-ph/0403021].
\bibitem{foot1} The set of nucleons
of energy 47 MeV produced by QGSJETII-3 that have a strange $\eta$
behavior (clearly visible in the upper right plot of Fig.
\ref{pseudorap1}) is assumed to be associated with a package
bug and not considered in our analysis.
\bibitem{Tar08} S.J. Sciutto and T. Tarutina, in preparation.
\bibitem{Sci01} S.J. Sciutto, in {\it Proceedings of the 27th ICRC, Hamburg, 2001}
(Copernicus Gesellschafr, Hamburg, 2001), Vol. 1, p. 237;
see also http://www.fisica.unlp.edu.ar/auger/aires.
\end{thebibliography}
\end{document}